\documentclass[apj]{emulateapj}

\usepackage{graphicx}
\usepackage{t1enc}
\usepackage[varg]{txfonts}
\usepackage{epsfig}
\usepackage{rotating}
\usepackage{natbib}
\usepackage{color}

\newcommand{\Trot}  {$T_\mathrm{rot}$}
\newcommand{\mum}   {$\mu$m}
\newcommand{\kms}   {km~s$^{-1}$}

\newcommand{\cmt}   {cm$^{-3}$}
\newcommand{\cmd}	{cm$^{-2}$}
\newcommand{\jpb}   {$\rm Jy~beam^{-1}$}    
\newcommand{\lo}    {$L_{\sun}$}

\newcommand{\mo}    {$M_{\sun}$}
\newcommand{\hh}	{H$_2$}

\newcommand{\nh}    {NH$_3$}

\newcommand{\hho} {H$_2$O}
\newcommand{\vel} {$v_\mathrm{LSR}$}
\newcommand{\velo} {$v$}
\newcommand{\et}    {et al.}
\newcommand{\eg}    {e.\,g.,}
\newcommand{\ie}    {i.\,e.,}

\newcommand{\uchii} {UC~H{\small II}}

\newcommand{\supa}  {$^\mathrm{a}$}
\newcommand{\supb}  {$^\mathrm{b}$}
\newcommand{\supc}  {$^\mathrm{c}$}
\newcommand{\supd}  {$^\mathrm{d}$}
\newcommand{\supe}  {$^\mathrm{e}$}
\newcommand{\supf}  {$^\mathrm{f}$}
\newcommand{\supg}  {$^\mathrm{g}$}

\definecolor{RED}{rgb}{1.0,0.0,0.0}

\shorttitle{NH$_3$ filaments in the IRDC G14.225--0.506}
\shortauthors{Busquet et al.}

\begin{document}

\title{Unveiling a network of parallel filaments in the Infrared Dark Cloud G14.225--0.506}

\author{Gemma Busquet\altaffilmark{1,2}, Qizhou Zhang\altaffilmark{3}, Aina Palau\altaffilmark{4}, Hauyu Baobab Liu\altaffilmark{5}, \'Alvaro S\'anchez-Monge\altaffilmark{6},
Robert Estalella\altaffilmark{2}, Paul T.~P. Ho\altaffilmark{3,5}, Itziar de Gregorio-Monsalvo\altaffilmark{7,8}, Thushara Pillai\altaffilmark{9}, Friedrich Wyrowski\altaffilmark{10}, Josep M. Girart\altaffilmark{4}, 
F\'abio P. Santos\altaffilmark{11}, and Gabriel A.~P. Franco\altaffilmark{11}}

\altaffiltext{1}{INAF-Istituto di Astrofisica e Planetologia Spaziali, via Fosso del Cavaliere 100, 00133 Roma, Italy}
\email{gemma.busquet@iaps.inaf.it}
\altaffiltext{2}{Departament d'Astronomia i Meteorologia, Institut de
Ci\`encies del Cosmos (ICC), Universitat de Barcelona (IEEC-UB), Mart\'i i Franqu\`es, 1,
E-08028 Barcelona, Catalunya, Spain}
\altaffiltext{3}{Harvard-Smithsonian Center for Astrophysics, 60 Garden Street, Cambridge, MA 02138, USA}
\altaffiltext{4}{Institut de Ci\`encies de l'Espai (CSIC-IEEC), Campus UAB, Facultat de Ci\`encies, Torre C-5 parell, E-08193 Bellaterra, Catalunya, Spain}
\altaffiltext{5}{Academia Sinica Institute of Astronomy and Astrophysics, Taipei, Taiwan}
\altaffiltext{6}{INAF, Osservatorio Astrofisico di Arcetri, Largo E.~Fermi 5, 05125 Firenze, Italy}
\altaffiltext{7}{European Southern Observatory, Karl-Schwarzschild-Strasse 2, 85748 Garching, Germany} 
\altaffiltext{8}{Join ALMA Observatory, Alonso de C\'ordova 3107, Vitacura, Santiago, Chile}
\altaffiltext{9}{Caltech Astronomy Department, MC 249-17, 1200 East California Boulevard, Pasadena, CA 91125, USA}
\altaffiltext{10}{Max-Planck-Institut f\"{u}r Radioastronomie, Auf dem H\"{u}gel 69, 53121 Bonn, Germany}
\altaffiltext{11}{Departamento de F\'isica-ICEx-UFMG, Caixa Postal 702, 30.123-970 Belo Horizonte-MG, Brazil} 

\begin{abstract}
We present the results of combined \nh\,(1,1) and (2,2) line emission observed with the Very Large Array and the Effelsberg 100~m telescope of the Infrared Dark Cloud G14.225--0.506. The \nh\ emission reveals a network of filaments constituting two hub-filament systems. Hubs are associated with gas of rotational temperature \Trot$\sim$15~K, non-thermal velocity dispersion $\sigma_{\mathrm{NT}}$$\sim$1~\kms, and exhibit signs of star formation, while filaments appear to be more quiescent (\Trot$\sim$11~K, $\sigma_\mathrm{NT}$$\sim$0.6~\kms). Filaments are parallel in projection and distributed mainly along two directions, at PA$\sim$10$\degr$ and 60$\degr$, and appear to be coherent in velocity. The averaged projected separation between adjacent filaments is between 0.5~pc and 1~pc, and the mean width of filaments is 0.12~pc. Cores within filaments are separated by $\sim$0.33$\pm$0.09~pc, which is consistent with the predicted fragmentation of an isothermal gas cylinder due to the `sausage'-type instability. The network of parallel filaments observed in G14.225--0.506 is consistent with the gravitational instability of a thin gas layer threaded by magnetic fields. Overall, our data suggest that magnetic fields might play an important role in the alignment of filaments, and polarization measurements in the entire cloud would lend further support to this scenario.
\end{abstract}

\keywords{stars: formation --- 
ISM: clouds ---
ISM: individual objects (G14.225--0.506)
}

\section{Introduction \label{sint}}

Filaments are ubiquitous structures in star forming complexes \citep[\eg][]{schneider1979, wiseman1998,hatchell2005,goldsmith2008,wang2008,jackson2010,schneider2010,molinari2010,andre2010,arzoumanian2011}, and often intersect in high-density regions of low aspect ratio and associated with star formation, known as hub-filament systems \citep[\eg][]{myers2009,liu2012}. However, 
their formation and their role in the star formation process is not well understood yet. 

In nearby ($d\sim$200--500~pc) molecular clouds, recent photometric results from \textit{Herschel} suggest that large-scale turbulence 
might be responsible for the formation of filaments \citep{arzoumanian2011}, while 
spectroscopic studies, sensitive to smaller scales, show that filaments present subsonic non-thermal motions \citep{hacar2011,pineda2011}, indicative of a dissipation of turbulence at smaller scales.
While this is consistent with observations in more distant and massive star-forming regions, such as G28.34+0.06 \citep{wang2008}, a number of studies reveal supersonic non-thermal motions and suggest the formation of filaments by the convergence of flows or by filament-filament collisions on large scales \citep{schneider2010,csengeri2011,heitsch2008,jimenez-serra2010,henshaw2012,miettinen2012,nakamura2012}. In addition, theoretical studies propose that magnetic fields could play a role in the formation of filaments \citep[\eg][]{nagai1998,nakamura2008}. It is clear, then, that several formation mechanisms 
have been invoked to explain the formation and alignment of filaments 
and therefore further spectroscopic studies of filamentary regions are essential to investigate 
the origin and evolution of such structures.

\begin{figure}[!t]
\begin{center}
\begin{tabular}[t]{c}
	\epsfig{file=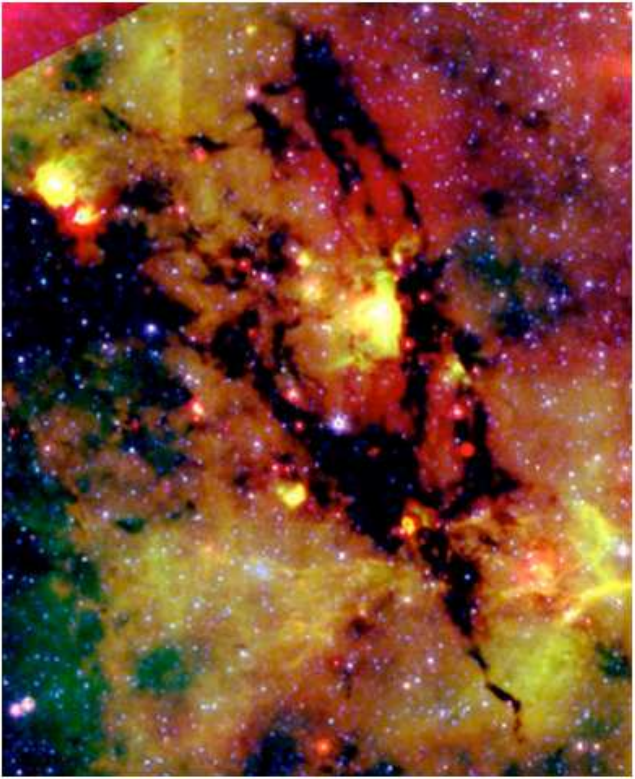,scale=0.85}	\\
	\end{tabular}
\caption{Archival \emph{Spitzer} 4.5/8.0/24~\mum\ (blue/green/red) three-color composite image of IRDC G14.225--0.506.
}
 \label{fg14spitzer}
\end{center}
\end{figure}

Filaments are prevailing structures in Infrared Dark Clouds (IRDCs, cf. \citealt{rathborne2006}). 
In this Letter we present combined Very Large Array (VLA) and Effelsberg\,100~m telescope observations of the \nh\,(1,1) and (2,2) transitions toward the IRDC\,G14.225--0.506 (hereafter G14.2).  Most of the studies performed so far toward this region focus on the brightest infrared sources, IRAS\,18153--1651 (hereafter I18153) and IRAS\,18152-1658 (hereafter I18152) with a luminosity of $\sim$1.1$\times10^4$~\lo\ and $\sim$4$\times10^{3}$~\lo, respectively, and located at a distance of 2.3~kpc \citep{jaffe1981,jaffe1982}.
Single-dish observations show that I18153 is associated with \hho\ maser emission \citep{jaffe1981,palagi1993}, and dense gas emission 
\citep{plume1992,anglada1996,bronfman1996}. More recent VLA observations reveal \hho\ maser emission in 9 different positions 
\citep{wang2006}, which indicates that star formation is already ongoing in some parts of the cloud. IRDC G14.2 
has been identified, using \emph{Spitzer} data, by  \citet{peretto2009} as a cloud containing an important amount ($\sim$100) of density enhancements or fragments displaying a filamentary morphology. 
The filamentary appearance of G14.2 (Fig.~\ref{fg14spitzer}) and its relatively nearby distance makes this region a good selection to investigate the physical properties of filaments and their formation mechanism
at high spatial resolution.

\section{Observations \label{sobs}}

The observations of the \nh\,(1,1) and (2,2) transitions were conducted using the VLA\footnote{
The Very Large Array (VLA) is operated by the National Radio 
Astronomy Observatory (NRAO), a facility of the National Science 
Foundation operated under cooperative agreement by Associated 
Universities, Inc.} in the D configuration on November 12 2005 (project AW666). We performed a 34-pointing mosaic covering an area of $7'\times13'$. The integration time was $\sim$4.5 minutes per pointing. The adopted flux density of the flux calibrator
3C\,286 was 2.41~Jy at a wavelength of 1.3~cm. The time variation of the gains was calibrated using J1832--105, with a bootstrapped flux of 0.97$\pm$0.01~Jy, and the bandpass calibrator used was 
3C\,273. We used the 4IF spectral line mode, which allows simultaneous observations of the \nh\,(1, 1) and (2, 2) lines with two 
polarizations for each line. The bandwidth used was 3.12~MHz, divided into 63 channels with a channel spacing of
48.8~kHz ($\sim$0.6~\kms\ at 1.3~cm), centered at $\sim$21~\kms. 
The visibility data sets were calibrated using the AIPS software package of the NRAO.

To recover extended structures filtered out by the interferometer, we performed  \nh\ observations with the Effelsberg 100~m telescope (project 101-07). The observations were carried out between April 4 and 7 2008. We used the 18-26~GHz HEMT receiver tuned to a frequency of 23.7~GHz with the 16384 channel Fast Fourier Transform Spectrometer, allowing simultaneous observations of the \nh\,(1,1) and (2,2) lines. The total bandwidth used was 100~MHz, which provides a velocity resolution of 0.075~\kms. The observations were conducted in frequency switching mode with a frequency throw of 7.5~MHz. 
At the observed wavelength, the half-power beamwidth  
of the telescope is $\sim$40$''$. The map covered an area of $8'\times13'$ and was made by observing the positions of a grid with half-beam spacing. Pointing was checked at hourly intervals, with a pointing accuracy better than 8$''$.
To convert the arbitrary noise tube units of the Effelsberg data to main beam brightness temperature we observed as a primary flux calibrator NGC\,7027 and a nearby quasar as a secondary flux calibrator. Data reduction was performed using the CLASS package, which is part of the GILDAS\footnote{http://www.iram.fr/IRAMFR/GILDAS} sofware.  
We combined the visibility data from the VLA and Effelsberg\,100~m telescope for both \nh\,(1, 1) and \nh\,(2, 2) lines following the MIRIAD procedure outlined in \citet{vogel1984}. 
We applied a $uv$-taper function of 23~$\mathrm{k}\lambda$ during imaging. The resulting synthesized beams were $8\farcs2\times7\farcs0$ (PA$=-15\degr$) for \nh\,(1,1) and 
$8\farcs0\times6\farcs9$ (PA$=-16\degr$) for \nh\,(2,2). The rms was $\sim$8~m\jpb\ per 0.6~\kms\ spectral channel.

\begin{figure*}[!ht]
\begin{center}
\begin{tabular}[t]{c}
	\epsfig{file=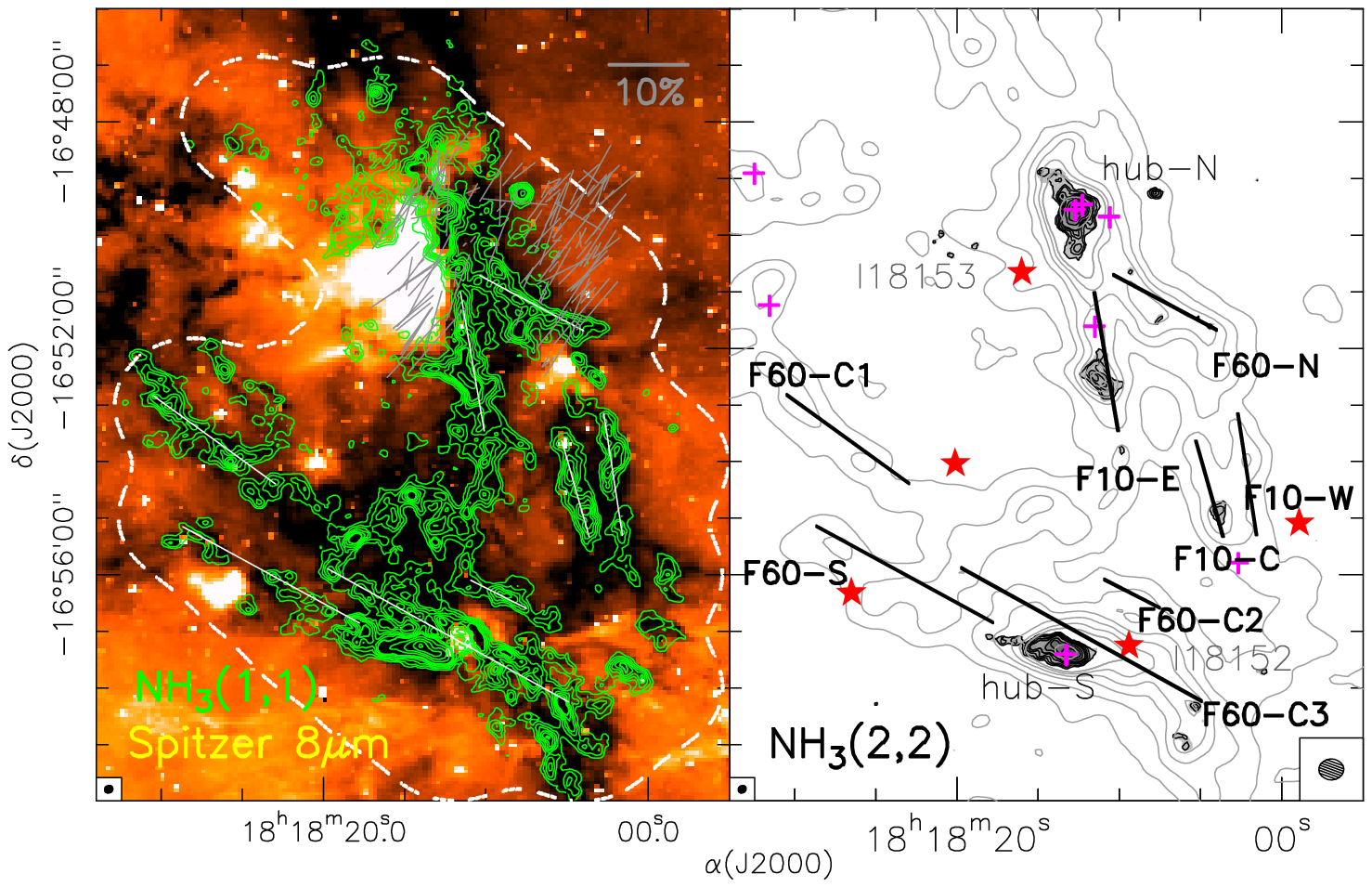,scale=1.1}	\\
		\epsfig{file=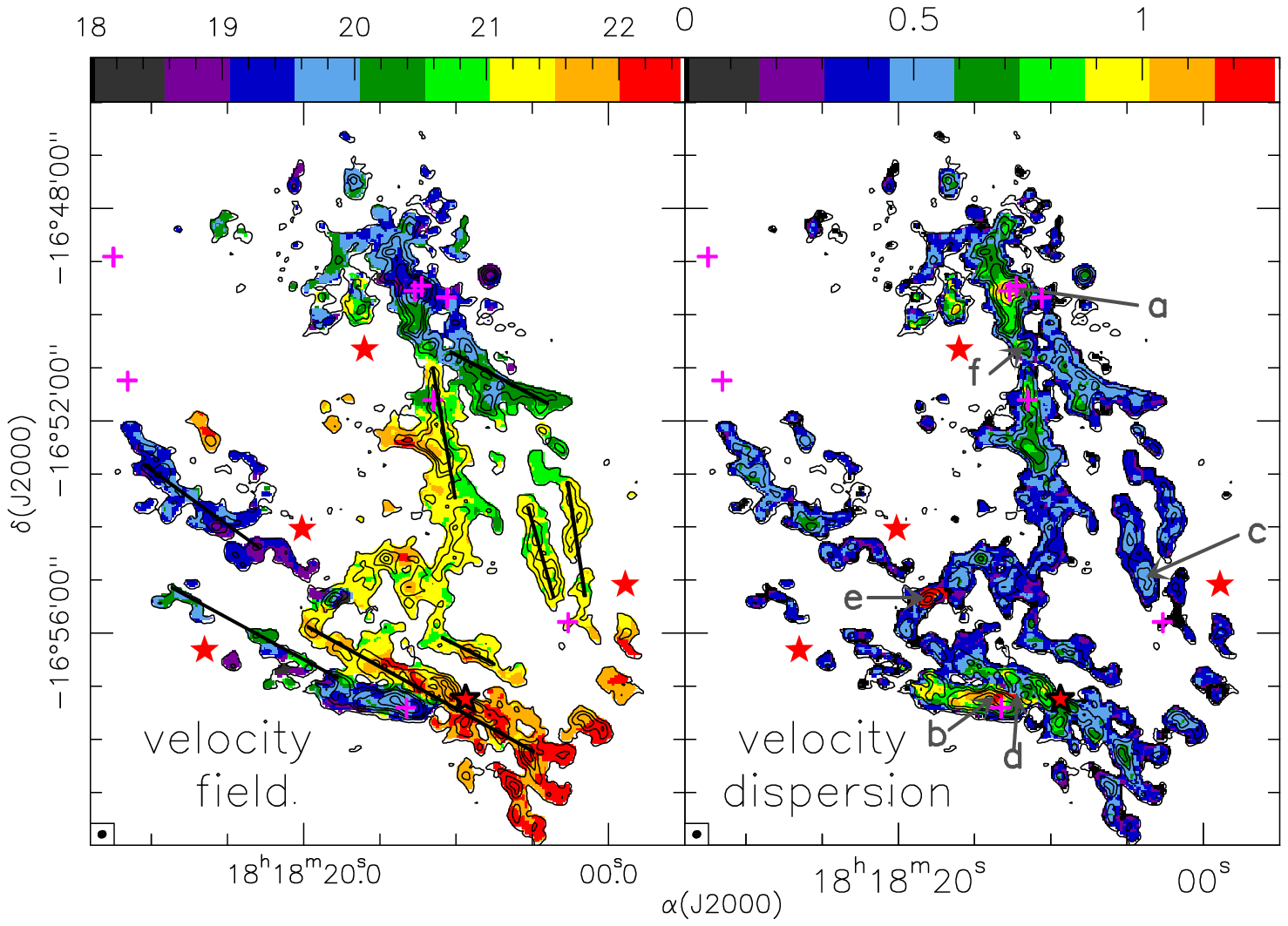,scale=1.1}	\
	\end{tabular}
\caption{\emph{Top Left:} Combined \nh\,(1,1) integrated intensity  (contours) overlaid on the 8~\mum\ \emph{Spitzer} image (color scale). Contour levels range from 3 to 18 in steps of 3, and from 18 to 58 in steps of 10 times the rms noise of the map, 9~m\jpb\,\kms. The dashed line indicates 50~\% of the sensitivity level of the VLA mosaic. Gray lines are the polarization vectors of the near-infrared ($H$-band) observations (Busquet \et\ in prep.) with the scale shown in the top right corner.
\emph{Top Right: } Combined \nh\,(2,2) integrated intensity (gray scale and black contours) overlaid on the 870~\mum\ continuum from LABOCA bolometer at the APEX telescope (gray contours; \citealt{busquet2010}). Contour levels for \nh\,(2,2) range from 2 to 10 in steps of 2, and from 10 to 60 in steps of 5 times the rms noise of the map, 9~m\jpb\,\kms. Contour levels for 870~\mum\ emission range from 3 to 53 in steps of 10, and from 53 to 653 in steps of 100 times the rms of the map, 25~m\jpb.
The \nh\ and 870~\mum\ continuum  synthesized beams are shown in the bottom left and bottom right corners, respectively. 
\emph{Bottom:} Contours: zero-order moment map of \nh\,(1,1). 
Color scale:  \nh\,(1,1) first-order moment map (\emph{Left}) and second-order moment map (\emph{Right}). Units are \kms. Stars indicate IRAS sources in the field, and crosses mark the position of \hho\ maser \citep{wang2006}. The most prominent filaments are indicated with white/black lines and labelled according to its position angle.
Arrows in the bottom right panel mark the positions of the \nh\,(1,1) spectra shown in Fig.~\ref{fg14spec}.
}
 \label{fg14nh3moms}
\end{center}
\end{figure*}

\section{Results and Analysis \label{sres}}

Figure~\ref{fg14nh3moms} (top left) shows the combined (VLA+Effelsberg) zero-order moment map of the \nh\,(1,1) emission 
overlaid on the 8~$\mu$m \emph{Spitzer} image. The overall morphology of the \nh\,(1,1) dense gas consists of extended and clumpy filamentary structures, strikingly mimicking the extinction feature seen in the \emph{Spitzer} image. 
While the \nh\,(1,1) emission is spatially extended, the \nh\,(2,2) emission is compact 
(Fig.~\ref{fg14nh3moms}-top right), suggesting that the extended emission is at lower temperatures. We identified the most prominent filaments based on the morphology of the \nh\,(1,1) together with the fact that these structures are coherent in velocity.  
We used the following criteria: \textit{i)} filaments must have aspect ratio larger than 6; \textit{ii)} the signal-to-noise ratio should be larger than 9\footnote{Signal-to-noise ratio computed in the zero-order moment map of \nh\,(1,1), where the rms noise level has been estimated using $3\sigma\Delta$\velo/$\sqrt{3}$, where $\sigma$ is the rms noise of the channel maps and $\Delta$\velo=0.6~\kms.}; and \textit{iii)} they must appear in at least two velocity channels and spanning a maximum velocity range of 3~\kms. 
Fig.~\ref{fg14nh3moms} (top right) shows, for comparison, the 870~\mum\ continuum emission from the LABOCA bolometer at the APEX telescope 
\citep{busquet2010}, supporting our identification. 

We identified a network of 8 filaments and two hubs (named hub-N and hub-S in Fig.~\ref{fg14nh3moms}), which were recognized using the \nh\,(2,2) emission as denser regions in which some filaments intersect. The \nh\ filaments, which cover a total area of 4.7$\times$8.7~pc, appear approximately parallel, in projection, in two preferred directions, at PA of 10$\degr$ and 60$\degr$, and they contain chains of dense cores\footnote{Cores have been identified in the zero-order moment map of \nh\,(1,1) if at least the $6\sigma$ level is closed, where $\sigma$ is the rms noise of the map.} aligned along the filament axis and distributed at somewhat regular spacings of about $\sim$30$''$ or 0.33$\pm$0.09~pc at the distance of the cloud. The averaged projected separation between adjacent filaments
is between 0.5~pc and 1~pc.
In Table~\ref{g14filaments} we report on the length and width at Full Width Half Maximum ($FWHM$) of each filament obtained from \nh\,(1,1) data. On average, we found that the  
aspect
ratio is $\sim$15:1, with a typical $FWHM$ width of $\sim$0.12~pc. This value is close to the filament width of 0.1~pc reported for the IC\,5146, Aquila, and Polaris molecular clouds from \textit{Herschel} observations \citep{arzoumanian2011}. 

\begin{figure}[!t]
\begin{center}
\begin{tabular}[t]{c}
	\epsfig{file=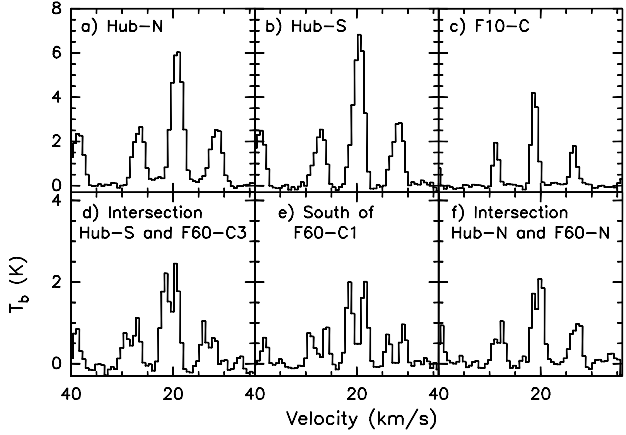,scale=0.47}	\\
	\end{tabular}
\caption{\nh\,(1,1) spectra, in units of brightness temperature, at some selected positions depicted with arrows in Fig.~\ref{fg14nh3moms}.}
\label{fg14spec}
\end{center}
\end{figure}

In Fig.~\ref{fg14nh3moms} (bottom left) we present the  first-order moment map 
of the \nh\,(1,1) main line. Within 
each filament the velocity variations are small, in the range of 1--2~\kms\ (see Table~\ref{g14filaments}), similar to other filamentary IRDCs \citep[\eg][]{jackson2010}. This network of filaments seems to be separated into two main velocity components, one at \vel$\sim$18.3--20.8~\kms\ and another one at \vel$\sim$20.8--23.2~\kms, which overlap in the hubs. 
The second-order moment map  
is presented in Fig.~\ref{fg14nh3moms} (bottom right), and shows that the velocity dispersion is locally enhanced ($\sigma$$\sim$1~\kms) toward hubs. Additionally, a high velocity dispersion ($\sigma$$\sim$1.6~\kms) is seen 
toward an arc-shaped structure connecting filament F10-E with the southern filaments, in a small region intersecting filament F60-C1 and labelled as position `e' in Fig.~\ref{fg14nh3moms}.
In this region
the large values of the velocity dispersion are due to the presence of two velocity components separated by $\sim$3~\kms\ (see Fig.~\ref{fg14spec}-e).
The presence of two velocity components is also found in regions where filaments intersect hubs (see Fig.~\ref{fg14spec}-d,f). 
In contrast, all the other filaments appear more quiescent, with a typical velocity dispersion of $\sim$0.4--0.6~\kms. 

\begin{table*}[!ht]
\begin{scriptsize}
\caption{Physical properties of filaments and hubs}
\centering
\begin{tabular}{l c c c c c c c c c c c c c}
\hline\hline\noalign{\smallskip}	
	&length\supa\ &width\supa\ &aspect &\vel\ &\Trot\supb~\supc\ &$\sigma_{\mathrm{TOT}}$\supc &$\sigma_{\mathrm{NT}}/c_{\mathrm{s}}$\supc\ &$M/l\,$\supc~\supd\ 	&$M_{\mathrm{vir}}/l\,$\supc~\supe\ &$\alpha_{\mathrm{vir}}$\supe\ &$\lambda_{\mathrm{obs}}$\supf\  &$\lambda_{\mathrm{f}}$\supf\  &$N_{\mathrm{cores}}$\supg\ \\
Filament &(pc) &(pc) &ratio &(\kms) &(K) &(\kms) & &(\mo\,pc$^{-1}$) &(\mo\,pc$^{-1}$) & &(pc) &(pc) &\\
\hline
\noalign{\smallskip}
F10-W        &1.48 &0.11    &13.5    	&20.8--22.0 	&10$\pm$2	&0.49	&2.38			&105	&112	&1.1	&0.27$\pm$0.08	&0.16--0.40	&\phn6\\
F10-C	 &1.18 &0.14   	&\phn8.4  &20.8--22.0 	&10$\pm$1	&0.50	&2.31			&147	&115	&0.8	&0.33$\pm$0.07	&0.16--0.41	&\phn4\\
F10-E	 &1.63 &0.15    &10.9       &20.2--22.0  	&16$\pm$1	&0.98	&3.63			&\phn87	&448	&5.1	&0.32$\pm$0.11	&0.21--0.80	&10\\
F60-N 	 &1.23 &0.09     &13.7        &19.5--20.8	&12$\pm$2 	&0.54	&2.31			&149	&135	&0.9	&0.25$\pm$0.09	&0.18--0.44	&\phn7\\
F60-C1	 &1.81  &0.12     &15.1        &18.3--20.2 	&10$\pm$3 	&0.63	&3.10			&120	&185	&1.5	&0.38$\pm$0.11	&0.16--0.52	&\phn8\\
F60-C2	&0.66   &0.09    &\phn7.3   &20.8--22.6	&10$\pm$3	&0.61	&3.03			&\phn74	&174	&2.3	&0.49$\pm$0.09  	&0.16--0.50      	&\phn5\\
F60-C3	&3.34  &0.12     &27.8        &21.4--23.2	&10$\pm$1	&0.73	&3.67			&218	&248	&1.1	&0.25$\pm$0.12	&0.16--0.60        &14\\
F60-S	&2.12  &0.10     &21.2    	&18.3--20.8  	&10$\pm$3 	&0.98	&5.02			&132	&447	&3.4	&0.38$\pm$0.11	&0.16--0.80	&\phn6\\
Hub-N	&1.12  &0.23	&\phn4.9  &17.7--20.8	&15$\pm$1	&0.98	&3.80			&266	&446	&1.7	&\ldots	&\ldots 		&\\
Hub-S	&1.15 &0.23  	&\phn5.0  &18.3--20.8	&15$\pm$1	&1.09	&4.31			&328	&548	&1.7	&\ldots\    &\ldots 		&\\
\hline
\end{tabular}
\begin{list}{}{}
\item[\supa]{Deconvolved size at Full Width Half Maximum ($FWHM$) not corrected for projection effects.}
\item[\supb]{\Trot\ has been derived following the appendix of \citet{busquet2009}. }
\item[\supc]{Averaged values within the area at $FWHM$. }
\item[\supd]{Mass per unit length, where the mass, $M=N$(\hh)\,2.8$m_{\mathrm{p}}\,A$, has been estimated assuming an \nh\ abundance of $3\times10^{-8}$ (average value  measured in IRDCs; Pillai \et\ 2006) and using the area $A$ of the filament at $FWHM$. The uncertainty in the mass is a factor of 3.}
\item[\supe]{Virial mass per unit length $M_{\mathrm{vir}}/l=2\sigma_{\mathrm{TOT}}^2/G$, and virial paramter $\alpha_{\mathrm{vir}}=M_{\mathrm{vir}}/M$ \citep{bertoldi1992}}
\item[\supf]{$\lambda_{\mathrm{obs}}$: observed separation between cores within a filament. $\lambda_{\mathrm{f}}$: 
predicted core separation $\lambda_{\mathrm{f}}=22\,H$, where $H=c_{\mathrm{s}}(4\pi\,G\,2.8m_{\mathrm{H}}\,n_{\mathrm{c}})^{-1/2}$ is the scale height, with $c_{\mathrm{s}}$ the isothermal sound speed (estimated by converting \Trot\ to kinetic temperature using the expression of \citet{tafalla2004}), $G$ the gravitational constant, and $n_{\mathrm{c}}$ the gas density at the center of the filament, adopted to be $10^5$~\cmt. The first value corresponds to the core separation using $c_{\mathrm{s}}$ and the second value was obtained replacing $c_{\mathrm{s}}$ by $\sigma_{\mathrm{TOT}}$.
\item[\supg]{Number of cores within each filament.}}
\end{list}
\label{g14filaments}
\end{scriptsize}
\end{table*}

To obtain the main physical properties (rotational temperature \Trot, total velocity dispersion $\sigma_{\mathrm{TOT}}$, and mass per unit length $M/l$) of each filament, we extracted an averaged spectrum of \nh\,(1,1) and (2,2) over the filament area at $FWHM$. The results are reported in Table~\ref{g14filaments}. 
The rotational temperature ranges between 10~K and 16~K.
The total velocity dispersion of the gas ranges from $\sim$0.5~\kms\ up to 1.1~\kms, and the non-thermal velocity dispersion over the isothermal sound speed, $\sigma_{\mathrm{NT}}/c_{s}$, ranges between 2 and 5, implying that filaments in G14.2 are characterized by supersonic non-thermal motions. 
In Table~\ref{g14filaments} we also list the mass and virial mass per unit length, the observed separation between cores, and the number of cores in each filament. Finally, 
the total surface density estimated by taking the spectrum averaged over all \nh\ filaments is $\Sigma$$\simeq$1.9~g\,\cmd.

\begin{figure}[t]
\begin{center}
\begin{tabular}[t]{c}
	\epsfig{file=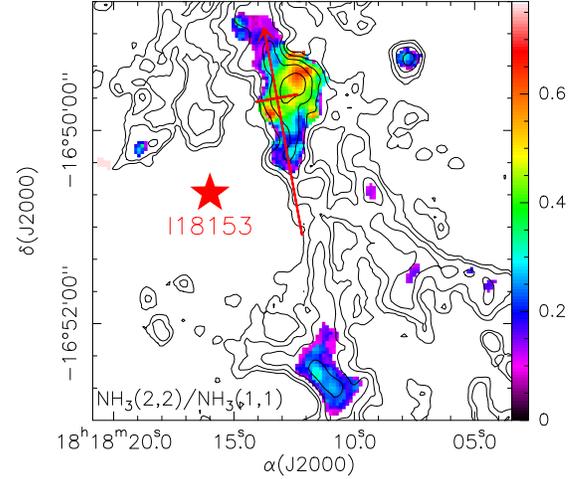,scale=0.6,angle=-90}\\
	\epsfig{file=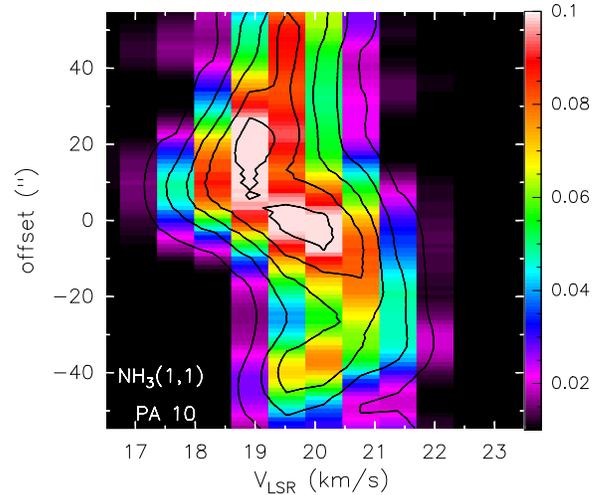,scale=0.6,angle=-90}	\\
	\end{tabular}
\caption{\emph{Top:} \nh\,(2,2)/\nh\,(1,1) map around I18153 (color scale) overlaid on the \nh\,(1,1) integrated intensity (contours). The long red arrow depicts the position-velocity (PV) cut, where the center of the cut is indicated by the intersection of the two red lines. \emph{Bottom:} PV plot of the \nh\,(1,1) emission along the cut at PA=10$\degr$. Contours start at 0.02 and increase in steps of 0.02~\jpb. Positive offsets increase as indicated by the arrow.}
\label{fg14pvplots}
\end{center}
\end{figure}

\section{Discussion and Conclusions}

In the previous section we presented the main properties of the two hub-filament systems in G14.225--0.506. 
Hubs are more compact (aspect ratio 5 vs 15), warmer (\Trot$\simeq$15~K vs 11~K), and show larger velocity dispersion and larger masses per unit length than filaments. 
Interestingly, hubs are associated with \hho\ maser emission (Wang \et\ 2006) and mid-infrared sources (see Figs.~\ref{fg14spitzer}, \ref{fg14nh3moms}),  and they are the main sites of stellar activity within the cloud. 

The stability of the filaments can be studied by estimating the virial parameter $\alpha_{\mathrm{vir}}$=$M_{\mathrm{vir}}/M$ \citep{bertoldi1992}, which is 
$\lesssim$2 for all the filaments and hubs except for F10-E, and five out of eight filaments are near virial equilibrium ($\alpha_{\mathrm{vir}}$$\simeq$1). This indicates that most of the filaments are unstable (collapsing) and probably undergoing fragmentation, compatible with the clumpy structure of G14.2. It is worth noting that filament F10-E has \Trot\ and velocity dispersion values similar to the hub properties.
Filament F10-E presents some striations converging toward it. However, while hub-N and hub-S seem to be places where two different velocity structures converge, F10-E shows only one velocity component. 
We speculate that I18153, an \uchii\ region with $L$$\sim$10$^4$~\lo\ (S\'anchez-Monge, private communication), may compress the gas, heating and injecting turbulence to this filament ($\alpha_{\mathrm{vir}}\simeq$5). The interaction of this \uchii\ region with the dense gas is also seen in hub-N, where the \nh\,(2,2)/\nh\,(1,1) map shows a local heating (Fig.~\ref{fg14pvplots}). The position-velocity (PV) plot along this hub (see Fig.~\ref{fg14pvplots}) reveals an inverted C-like structure, consistent with an expanding shell \citep{arce2011}. 

We investigated the fragmentation of filaments in the magnetohydrodynamic `sausage'-type instability scenario (Chandrasekhar \& Fermi 1953, see also Jackson \et\ 2010),
which predicts periodic separation between fragments (or cores) for a given density and isothermal sound speed. For an isothermal gas cylinder of finite radius $R$, the core separation can be expressed as  $\lambda_{\mathrm{f}}$=22$H$ for $R$$\gg\,$$H$, where $H$ is the scale height (see Table~\ref{g14filaments} for the formal expression). 
This is the case of G14.2, since $R$ and $H$ are 0.12~pc and 0.04~pc, respectively. Adopting a density of $10^5$~\cmt, we estimated the predicted core separation using first the isothermal sound speed, yielding $\lambda_{\mathrm{f}}$$\sim$0.16--0.21~pc, and then replacing $c_{\mathrm{s}}$ by the total velocity dispersion 
$\sigma_{\mathrm{TOT}}$, which gives $\lambda_{\mathrm{f}}$$\sim$0.4--0.8~pc (see Table~\ref{g14filaments}). 
The observed separation, $\sim$0.33~pc, is in agreement with these two extreme cases. It is noteworthy that most of the cores appear to be elongated along the direction of the filament, which could imply the possibility of further fragmentation at smaller scales as observed in the IRDC\,G28.34+0.06-P1 \citep{zhang2009,wang2011}. 

One of the most intriguing features of G14.2 is the network of filaments that are aligned in parallel. The filaments appear to take two
preferred directions, one group at a PA of 10$\degr$, and the others at a PA of 60$\degr$. This network of filaments  
may arise from a layer of
self-gravitating gas. Instability analysis has been performed for a layer
of an isothermal infinite sheet under dynamical perturbation \citep{ledoux1951,schmid-burgk1967,elmegreen1978,larson1985,nagai1998,myers2009}. The gas
is unstable to perturbations, which leads to high density columns in the plane as a result
of gravitational instability. The spacings between the high density columns of gas correspond
to the wavelength of the fastest growth mode. In the absence of magnetic fields, the growth
of instability does not have a preferred direction in the plane. As a result, a grid of connected filaments
appears in the gas layer. If the gas layer is threaded by magnetic fields, the growth of the instability
develops unrestricted in one direction and is suppressed along the orthogonal direction. 
\citet{nagai1998} analyzed a pressure confined isothermal gas layer threaded
by uniform magnetic fields. They found that in the regime of smaller external pressure (\ie\
the scale height $H$$\ll$$Z_{\mathrm{b}}$, where $Z_{\mathrm{b}}$ is the thickness of the gas layer) the instability grows faster along the field lines.
As a consequence, high density columns, or filaments develop with their longitudinal axis
perpendicular to the field lines. In the high pressure regime (\ie\ $H$$\gg$$Z_{\mathrm{b}}$), the fastest
growth of instability is perpendicular to the field lines and gives rise to filaments
parallel to the magnetic fields.

Recent numerical simulations (Van Loo, private communication) confirm the linear analysis
in \citet{nagai1998}. The simulations show that an array of high density columns develop in
the gas layer with magnetic fields. In addition, lower density filamentary structures are also
present, inter-connecting the main filaments. The highest density structures are found at the intersections
of major and minor filaments, as in this work. Furthermore, the simulations show that gravitational instability
develops within a filament during the filament formation. This means that fragmentation of a filament into cores
occurs simultaneously with the fragmentation of the sheet, but according to \citet{toala2012} with different free-fall times.

The array of filamentary structures in G14.225--0.506 may arise from gravitational
instability of a thin gas layer with magnetic fields. In fact, preliminary near-infrared polarimetric observations around hub-N (Busquet \et\ in prep.) reveal that the magnetic field is perpendicular to filaments at PA$\sim$60$\degr$
(see Fig.~\ref{fg14nh3moms}). Therefore, according to \citet{nagai1998} G14.2 would be in the regime of small external pressure ($H\ll\,Z_{\mathrm{b}}$). 
Using the total surface density (see Sect.~3), the scale height $H$ of the initial gas layer is $H\sim$0.09~pc.
This value should be regarded with caution, and to definitively assess its validity one would need observations of a low-density gas tracer to be sensitive to the gas layer.  
The wavelength of the fastest mode can be expressed as $\lambda_{\mathrm{fastest}}=4\pi\,H$ (Eq.~60 in \citealt{nagai1998}). Using our estimation of $H$, the predicted separation is $\sim$1.1~pc, in 
agreement with the observed filament separation (between 0.5 and 1~pc). 
It is not clear how such a large gas layer (4.7$\times$8.7~pc) may form initially.
The convergence of dynamic flows could be responsible for the formation of such a large gas layer that subsequently could fragment into parallel filaments as a result of magnetic modulation. Our \nh\ data, although showing two velocity components,  
do not reveal evidence of converging/interacting flows and further observations of low-velocity shock tracers, like SiO or CH$_3$CN 
\citep{jimenez-serra2010,csengeri2011}, and a tracer of low density material are required to definitely identify signatures of converging flows. 
Overall, our data suggest that magnetic fields might play an important role in the alignment of filaments, and polarization measurements in the entire cloud would lend further support to this scenario.


\acknowledgments
\begin{small}

The authors are grateful to the anonymous referee for valuable comments. G.B. is deeply grateful to Eugenio Schisano for very fruitful discussion on filaments. 
G.B is funded by an 
Italian Space Agency (ASI) fellowship under contract number I/005/07/0. A.P., R.E., and I.d.G. are  supported by the Spanish MICINN grant AYA2011-30228-C03 (co-funded with FEDER funds). A.~P. is supported by a JAE-Doc CSIC fellowship co-funded with the European Social Fund, under the program `Junta para la Ampliaci\'on de Estudios', and by the AGAUR grant 2009SGR1172 (Catalonia). F.P.S. and G.A.P.F. are partially supported by CNPq and FAPEMIG. This work is partially based on observations with the 100~m telescope of the MPIfR (Max-Planck-Institut f{\"u}r Radioastronomie) at Effelsberg.

\end{small}

\bibliographystyle{apj} 


\end{document}